\documentclass[vecphys]{svmult}
\usepackage{graphicx}
\usepackage{natbib}
\usepackage{multicol}        % used for the two-column index
\usepackage[bottom]{footmisc}% places footnotes at page bottom

\def\arcsec{\hbox{$^{\prime\prime}$}}
\newcommand{\Afr}{Af\rho{ }}
\begin{document}
\title*{Observations of Comet C/LINEAR (2004B1) between
2 and 3 AU heliocentric distance
\thanks{Based on observations obtained in the National Astronomical
Observatory, Rozhen}
}
\titlerunning{Comet C/LINEAR (2004B1) between 2 and 3 AU heliocentric
distance}
\author{
 Tanyu Bonev\inst{1}\and
 Galin Borisov\inst{1}\and
 Andon Kostov\inst{1}\and
 Abedin Abedin\inst{1}\and
 Gordana Apostolovska\inst{3}\and
 Bonka Bilkina\inst{1}\and
 Zahary Donchev\inst{1}\and
 Violeta Ivanova\inst{1}\and
 Vladimir Krumov\inst{1,2}\and
 Yanko Nikolov\inst{1,2}\and
 Evgeni Ovcharov\inst{2}\and
 Vesselka Radeva\inst{1}\and
 Ivaylo Stanev\inst{2}}
\authorrunning{Tanyu Bonev, Galin Borisov, Andon Kostov et al.}
\institute{
Institute of Astronomy and National Astronomical Observatory, \\
           Bulgarian Academy of Sciences, \texttt{tbonev@astro.bas.bg}
\and
Department of Astronomy, Sofia University "St. Kliment Ohridski"\
\and
Institute of Physics, Faculty of Natural Sciences, Skopje, Macedonia
}

\maketitle
\setcounter{footnote}{0}
\begin{abstract}
We present R-band observations of comet C2004B1 obtained in the period June, 21 -
August 20, 2006. The data have been reduced to surface brightness maps, light curves,
and  mean radial profiles of the coma. In two cases a decrease of the brightness was recorded, 
which lasted for several days. The brightness decrease was accompanied by
morphological changes in the coma. 
\end{abstract}

{\it Keywords:} comet 2004B1, coma brightness

\section{Introduction}
On January 29.16 the Lincoln Laboratory Near-Earth Asteroid Research project
(LINEAR) discovered a faint distant comet. The new comet was named C/LINEAR
(2004B1).  For shortness, we will refer to it C2004B1. The orbit of the
comet is slightly hyperbolic (eccentricity = 1.001), and its perihelion was
on 2006, Feb 7.9.  We present images of the comet, obtained during a
monitoring campaign lasting for 2 months, and point to correlations between
morphological changes in the coma and variations of the total brightness

%%%%%%%%%%%%%%%%%%%%%%%%%%%%%%%%%%
\section{Observations}
We observed the comet on its outbound orbit, between 2.4 and 2.9 AU
heliocentric distance. 
Our observing campaign started on June 21, 2006, and lasted until August, 20. 
The observations were carried out with the Schmidt telescope of the National
Astronomical Observatory. The images were obtained in the R-band with a CCD camera ST-8. Parameters of the camera can be seen in Kostov et al. \cite{AK:07}.
Details for the observations are given in table 1. Visit
http://www.astro.bas.bg/solsys/C2004B1/ to see all the images.
\begin{center}
\begin{table}[h]
\caption{Data for the observations}
\vspace{5mm}
\begin{center}
\begin{tabular}{crcccc}
\hline
Month   & Day & Days after & heliocentric & geocentric & Phase \\
of 2006 &     & perihelion & distance     & distance   & angle \\
\hline \hline
  June &  21 &    133 & 2.363 & 1.650  & 21.1 \\
       &  22 &    134 & 2.372 & 1.668  & 21.2 \\
       &  29 &    141 & 2.436 & 1.803  & 21.9 \\
  July &   7 &    149 & 2.509 & 1.974  & 22.4 \\
   &   8 &    150 & 2.519 & 1.996  & 22.4 \\
   &   9 &    151 & 2.528 & 2.018  & 22.5 \\
   &  13 &    155 & 2.565 & 2.109  & 22.5 \\
   &  14 &    156 & 2.574 & 2.132  & 22.5 \\
   &  16 &    158 & 2.593 & 2.178  & 22.5 \\
   &  27 &    169 & 2.697 & 2.438  & 22.1 \\
   &  29 &    171 & 2.715 & 2.486  & 21.9 \\
  August &   2 &    175 & 2.753 & 2.581  & 21.6 \\
   &   3 &    176 & 2.763 & 2.604  & 21.5 \\
   &   4 &    177 & 2.772 & 2.628  & 21.4 \\
   &   5 &    178 & 2.782 & 2.652  & 21.3 \\
   &   9 &    182 & 2.820 & 2.746  & 20.9 \\
   &  11 &    184 & 2.839 & 2.793  & 20.7 \\
   &  12 &    185 & 2.849 & 2.816  & 20.6 \\
   &  13 &    186 & 2.858 & 2.840  & 20.5 \\
   &  14 &    187 & 2.868 & 2.863  & 20.4 \\
   &  15 &    188 & 2.877 & 2.886  & 20.2 \\
   &  16 &    189 & 2.887 & 2.909  & 20.1 \\
   &  17 &    190 & 2.896 & 2.932  & 20.0 \\
   &  18 &    191 & 2.906 & 2.955  & 19.9 \\
   &  19 &    192 & 2.915 & 2.978  & 19.7 \\
   &  20 &    193 & 2.925 & 3.001  & 19.6 \\
\hline
\end{tabular}
\end{center}
\end{table}
\end{center}
%%%%%%%%%%%%%%%%%%%%%%%%%%%%%%%%%%%%%%%%%%%%5
\section{Surface brightness distribution}
\begin{figure}[h]
\centering
\includegraphics[width=\textwidth]
{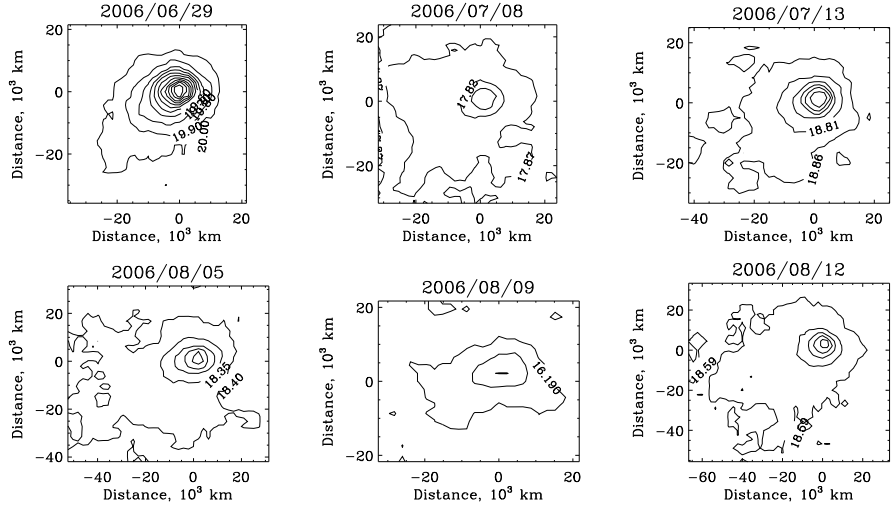}
\caption{Surface brightness of comet C2004B1 on 6 selected days. 
The isophote levels are in magnitudes/(arcsec)$^2$.
Note, in both rows the morphology of the coma in the middle panel is different
compared to the left and right images (see the text for more information). 
}
\label{fig:maps}
\end{figure}

The images were converted to surface brightness by using
standard stars in the field. We used the R-magnitudes of these stars given
in the USNO (United States Naval Observatory)\footnote{http://www.usno.navy.mil/} catalogue for  photometric calibration. 
% The relatively low photometric
% accuracy of this catalogue was compensated statistically by using a large  $$$$$$$$ number of standard stars. 
Examples of surface brightness maps are shown in figure \ref{fig:maps}. 
The 6 images shown are not a randomly selected sample. 
Their purpose is to illustrate how the coma morphology 
changes with variations of the total comet magnitude. 
The maps in the middle panels are derived from the images obtained at the two instants of reduced brightness, described in the next section. 
%%%%%%%%%%%%%%%%%%%%%%%%%%%%%%%%%%%%%%%%%
\section{Light curves}
\label{sec:light}
\begin{figure}[h]
\centering
\includegraphics[bb = 54 16 505 503,  width=0.30\textwidth, clip=]
{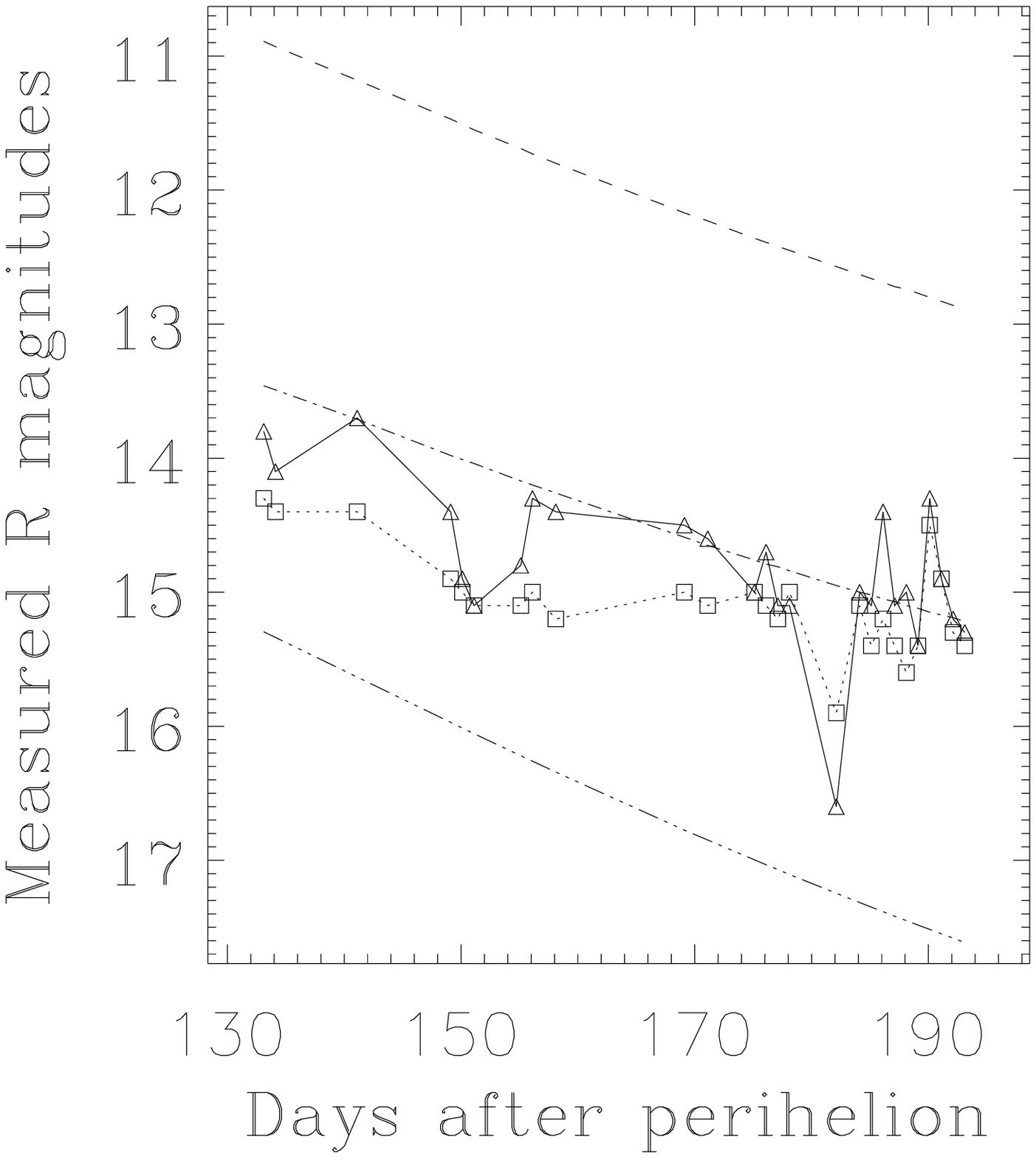}
\hfill
\includegraphics[bb = 54 16 505 503,  width=0.31\textwidth, clip=]
{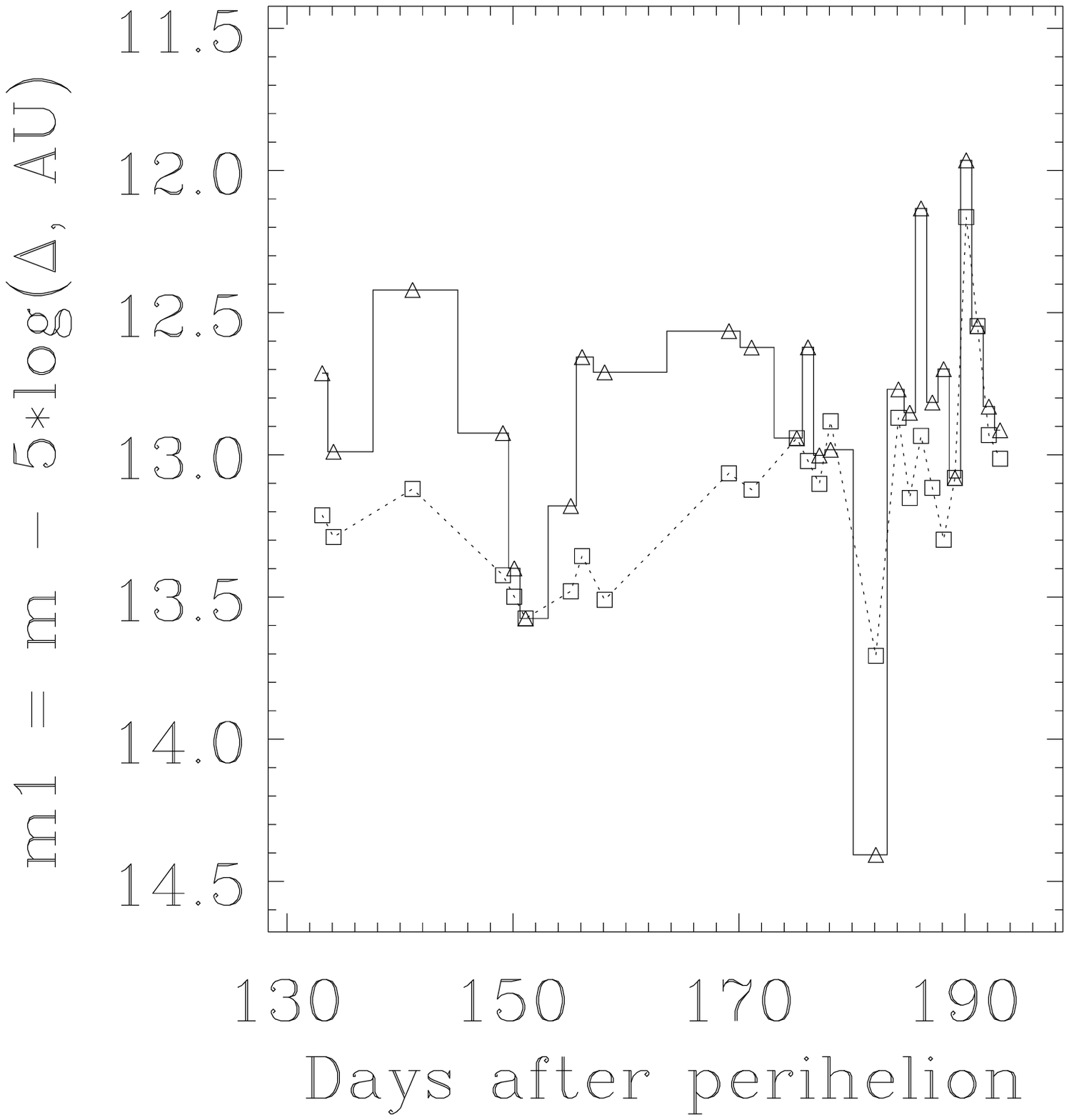}
\hfill
\includegraphics[bb = 54 16 505 503,  width=0.31\textwidth, clip=]
{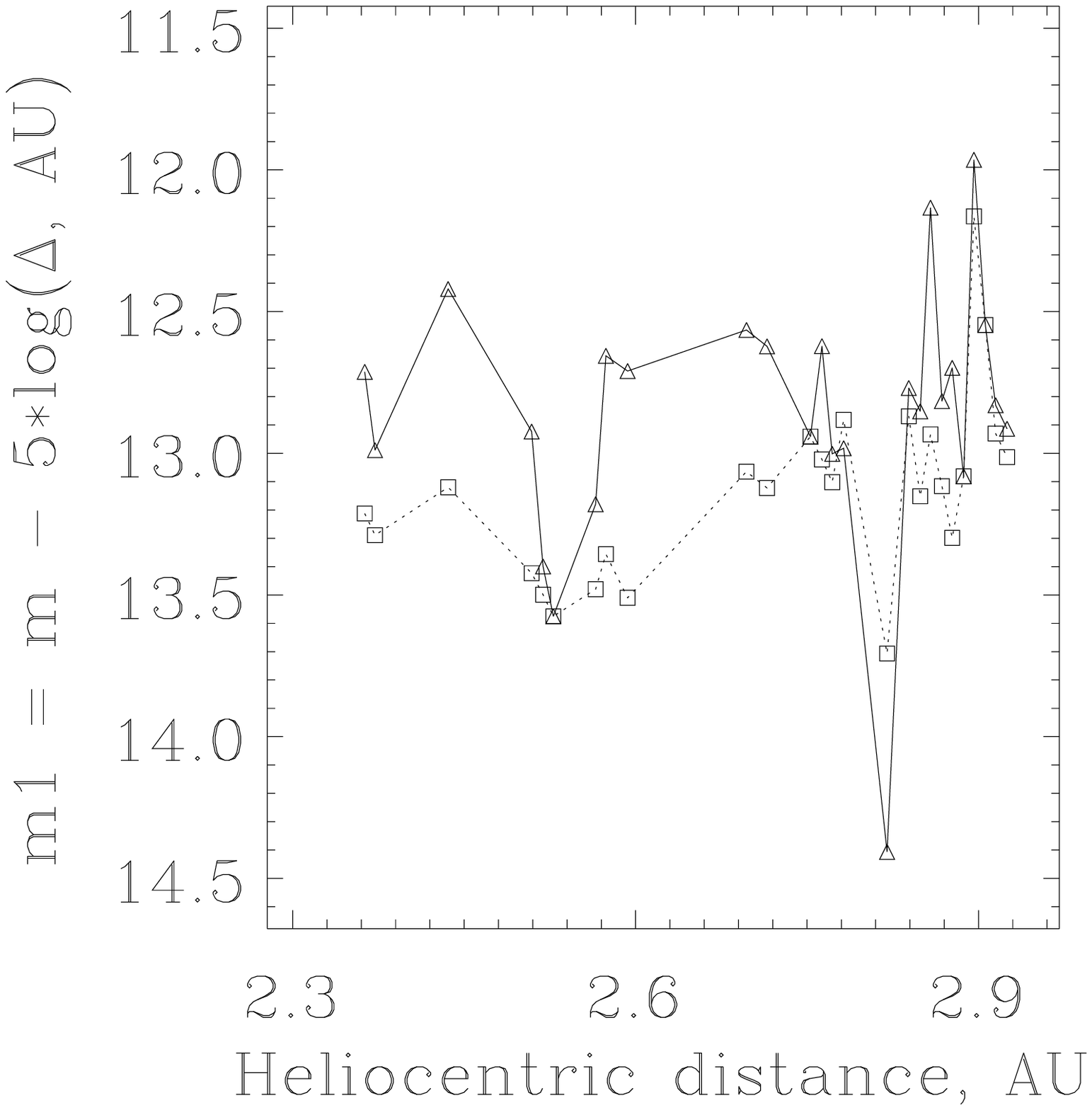}
\caption{Left: R magnitudes against time.
Triangles and full line: total brightness. 
Squares and dotted line: Brightness in a circular aperture of radius 10\arcsec. 
Dashed line: Total brightness from old ephemeris (Nov 8, 2006). 
Dash-dot line: Brightness of the nucleus from this ephemeris.  
Dash-double dot line: Total brightness in new ephemeris (April 1, 2007). 
Middle: R magnitudes, corrected for geocentric distance, against time. 
Right: R magnitudes, corrected for geocentric distance, against 
heliocentric distance. 
}
\label{fig:fig1}
\end{figure}

Figure \ref{fig:fig1} shows the light curves of comet C2004B1. They are compared to ephemeris calculated at two different epochs by HORIZONS\footnote{ http://ssd.jpl.nasa.gov/horizons.cgi}. The first one (dotted line) is from Nov 8, 2006 and is based on 1521 observations from the period 2004 - 2006. This ephemeris gives also the brightness of the comet nucleus, shown with dashed-dotted line in figure \ref{fig:fig1}. The second ephemeris data (dashed-double dot line) in the same figure, are from April 1, 2007, and are based on 1576 observations from the period 2004 - 2007. 
Five months after perihelion the brightness of comet C2004B1 was
substantially reduced in comparison to the first
ephemeris, but about one magnitude above the values from the second one.  
In most cases the total brightness of the comet was comparable
with the predicted brightness of the bare nucleus given by the first ephemeris.
In two cases, on July 7,8,9 and around Aug 9, the brightness of
the comet dropped more than 1 magnitude below the mean level. 
%%%%%%%%%%%%%%%%%%%%%%%%%%%%%%%%%%%%%
\section{Discussion}
Figure \ref{fig:fig1} (right panel) shows that the activity of the comet is
almost independent on its heliocentric distance. In their taxonomic study of
85 comets A'Hearn et al. \cite{A'H:95} found cases in which the dust
production is even increasing with heliocentric distance. According to these
authors flat slopes are characteristic for dynamically new comets before
their perihelion passage. We remind that comet C2004B1 was observed several
months after its perihelion.

\begin{figure}[h]
\includegraphics[bb = 35 15 509 396, width=0.46\textwidth, clip=]
{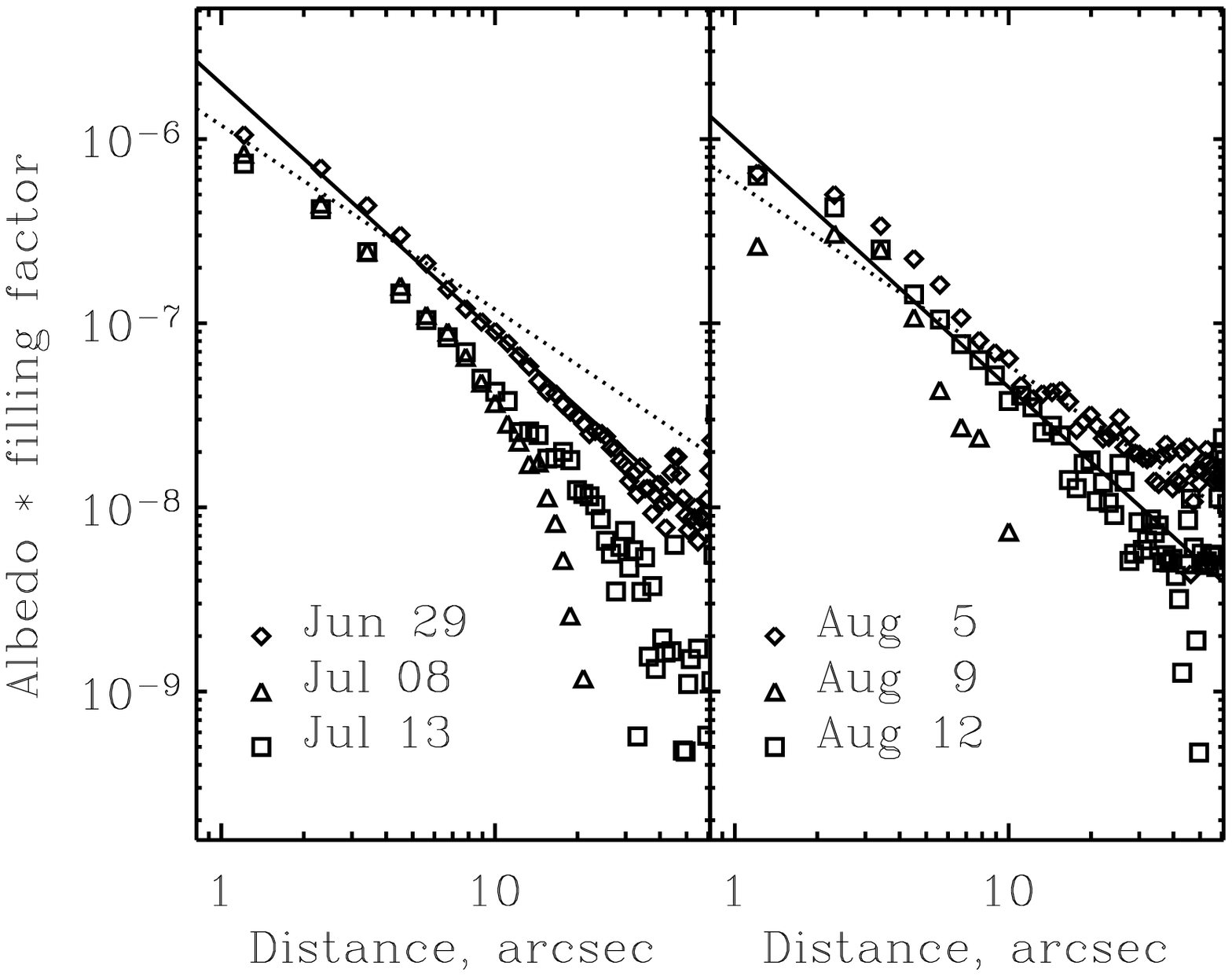}
\hfill
\includegraphics[bb = 35 15 509 396, width=0.46\textwidth, clip=]
{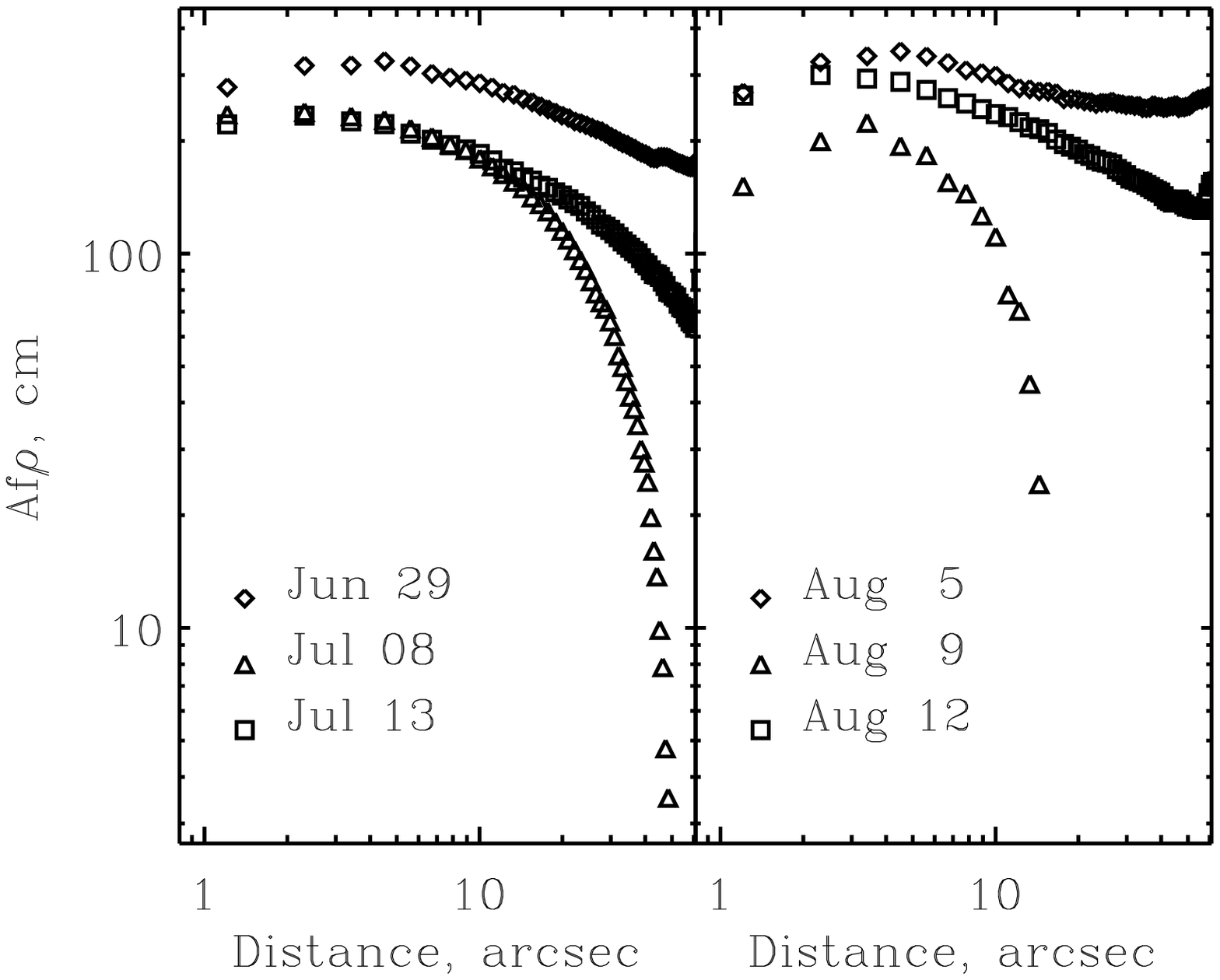}
\caption{Left: The two panels show azimuthally averaged profiles of the images
obtained around periods of reduced brightness. Triangles denote the data at
the minima, diamonds and squares stand for the days of normal activity before
and after the minima. 
Right: The corresponding $\Afr$ profiles.
}
\label{fig:mean_prof}
\end{figure}

The decreased total brightness of the comet at two occasions (see figure
\ref{fig:fig1}) is accompanied by morphological changes in the coma.  Most
remarkable is the faster decrease of the surface brightness in the outer
coma, at distances $>$ 10\arcsec.  This is well seen in figure
\ref{fig:mean_prof}, where in the left panel 
the azimuthally averaged profiles of the images with reduced brightness are
compared to those obtained several days before the brightness decrease and
several days after that.  The dotted line shows a $\rho^{-1}$-profile, the
ideal case of an isotropic and stationary outflow.  The full line is a
$\rho^{-1.35}$ - law which fits well the data obtained during phases of
normal activity. This steeper slope of the mean brightness profile is most
probably due to the gradual increase of the velocity with increasing
cometocentric distance, caused by the accumulated influence of the radiation
pressure acceleration. In the right panel of the same figure the
corresponding $\Afr$ profiles are shown ($\Afr$ was introduced by A'Hearn et
al. \cite{A'H:84} as a proxy for the dust production rate).  Comparison of
the profile from July 8 (drop of brightness) with the profile from July 13
(almost return to the total magnitude before the drop) shows comparable
levels of the mean surface brightness at distances $<$ 10\arcsec. This is an
indication that most of the brightness reduction is due to losses in the
outer coma, rather than to reduced activity of the nucleus. In the August
event the reduction is again stronger in the outer coma but now it is
accompanied by decrease in the circumnuclear region. In this case both,
reduction of the dust production from the nucleus and losses in the coma
should be responsible for the total brightness decrease. The reduced
brightness of the outer coma could be explained by sublimation of dust
particles. Recently, Beer et al.
\cite{Beer:06} have shown how sublimation of dirty icy grains can influence
the particle size distribution in the coma and lead to a dominant role of
large particles.
%%%%%%%%%%%%%%%

{\bf Do large dust grains dominate the coma of comet C2004B1 on its outbound
orbit between 2.4 AU and 2.9 AU perihelion distance?  What are the reasons
for the sudden drop of brightness?  Answers to these questions could be
given by comparison of our data with infrared observations}

{\em Acknowledgments:} This work made extensive use of the JPL Horizons
On-Line Ephemeris System (Giorgini et al., \cite{Horizons}). The Online
USNO-A2.0 Catalogue Server at the ESO/ST-ECF
Archive\footnote{http://archive.eso.org/skycat/servers/usnoa} was used to
extract data from the USNO catalogue.

\end{document}